\documentclass[longbibliography,aps,reprint,groupedaddress,superscriptaddress]{revtex4-1}
\usepackage{lineno}
\usepackage{graphicx}
\usepackage{float}
\usepackage{amssymb,amsmath}
\usepackage{mathrsfs}
\usepackage[T1]{fontenc}
\usepackage{color}

\newcommand{\rfig}[1]{Fig.~\ref{#1}}
\newcommand{\rfigs}[1]{Figs.~\ref{#1}}
\newcommand{\req}[1]{Eq.~(\ref{#1})}

\definecolor{blueprl}{rgb}{0.157, 0.173, 0.569}
\usepackage{hyperref}
\hypersetup{colorlinks=true, linkcolor=blueprl, citecolor=blueprl, urlcolor=blueprl,anchorcolor=blueprl}

\newenvironment{shrinkeq}[1]
{ \bgroup
	\addtolength\abovedisplayshortskip{#1}
	\addtolength\abovedisplayskip{#1}
	\addtolength\belowdisplayshortskip{#1}
	\addtolength\belowdisplayskip{#1}}
{\egroup\ignorespacesafterend}

\newcommand{\XX}{X\!X}
\newcommand{\YY}{Y\!Y}
\newcommand{\ZZ}{Z\!Z}

\newenvironment{sequation}{\begin{equation}\small}{\end{equation}}
\newenvironment{seqnarray}{\begin{eqnarray}\small}{\end{eqnarray}}

\DeclareMathSizes{10}{9}{7}{5}

\newcommand{\SIQSE}{\affiliation{1}{Shenzhen Institute for Quantum Science and Engineering, Southern University of Science and Technology, Shenzhen, Guangdong, China}}
\newcommand{\DPHY}{\affiliation{2}{Department of Physics, Southern University of Science and Technology, Shenzhen, Guangdong, China}}
\newcommand{\IQA}{\affiliation{3}{International Quantum Academy, Shenzhen, Guangdong, China}}
\newcommand{\GDKL}{\affiliation{4}{Guangdong Provincial Key Laboratory of Quantum Science and Engineering, Southern University of Science and Technology, Shenzhen, Guangdong, China}}
\newcommand{\HFNL}{\affiliation{5}{
		Shenzhen Branch, Hefei National Laboratory, Shenzhen 518048, China}}

\begin{document}
	\title{Floquet Engineering of Anisotropic Transverse Interactions in Superconducting Qubits} 
	\author{Yongqi Liang}
	\thanks{These authors contributed equally to this work.}
	\affiliation{\SIQSE}\affiliation{\IQA}\affiliation{\GDKL}
	
	\author{Wenhui Huang}
	\thanks{These authors contributed equally to this work.}
	\affiliation{\SIQSE}\affiliation{\IQA}\affiliation{\GDKL}
	
	\author{Libo Zhang}
	\thanks{These authors contributed equally to this work.}
	\affiliation{\SIQSE}\affiliation{\IQA}\affiliation{\GDKL}
	
	\author{Ziyu Tao}
	\affiliation{\SIQSE}\affiliation{\IQA}\affiliation{\GDKL}\affiliation{\DPHY}
	
	\author{Kai Tang}
	\affiliation{\DPHY}
	
	\author{Ji Chu}
	\affiliation{\IQA}
	
	\author{Jiawei Qiu}
	\affiliation{\SIQSE}\affiliation{\IQA}\affiliation{\GDKL}
	
	\author{Xuandong Sun}
	\affiliation{\SIQSE}\affiliation{\IQA}\affiliation{\GDKL}\affiliation{\DPHY}
	
	\author{Yuxuan Zhou}
	\affiliation{\IQA}
	
	\author{Jiawei Zhang}
	\affiliation{\SIQSE}\affiliation{\IQA}\affiliation{\GDKL}
	
	\author{Jiajian Zhang}
	\affiliation{\SIQSE}\affiliation{\IQA}\affiliation{\GDKL}\affiliation{\DPHY}
	
	\author{Weijie Guo}
	\affiliation{\IQA}
	
	\author{Yang Liu}
	\affiliation{\IQA}
	
	\author{Yuanzhen Chen}
	\affiliation{\SIQSE}\affiliation{\IQA}\affiliation{\GDKL}\affiliation{\DPHY}
	
	\author{Song Liu}
	\affiliation{\SIQSE}\affiliation{\IQA}\affiliation{\GDKL}\affiliation{\HFNL}
	
	\author{Youpeng Zhong}
	\email{zhongyp@sustech.edu.cn}
	\affiliation{\SIQSE}\affiliation{\IQA}\affiliation{\GDKL}\affiliation{\HFNL}
	
	\author{Jingjing Niu}
	\email{niujj@iqasz.cn}
	\affiliation{\IQA}\affiliation{\HFNL}
	
	\author{Dapeng Yu}
	\affiliation{\SIQSE}\affiliation{\IQA}\affiliation{\GDKL}\affiliation{\DPHY}\affiliation{\HFNL}

	\begin{abstract}
		Superconducting transmon qubits have established as a leading candidate for quantum computation, as well as a flexible platform for exploring exotic quantum phases and dynamics.
		However, physical coupling naturally yields isotropic transverse interactions between qubits, restricting their access to diverse quantum phases that require spatially dependent interactions. Here, we demonstrate the  simultaneous realization of both pairing ($\XX-\YY$) and hopping ($\XX+\YY$) interactions between transmon qubits by Floquet engineering. The coherent superposition of these interactions enables independent control over the $\XX$ and $\YY$ terms, yielding  anisotropic transverse interactions. By aligning the transverse interactions along a 1D chain of six qubits, as calibrated via Aharonov-Bohm interference in synthetic space, we synthesize a transverse field Ising chain model and explore its dynamical phase transition under varying external field. The scalable synthesis of anisotropic transverse interactions paves the way for the implementation of more complex physical systems requiring spatially dependent interactions, enriching the toolbox for engineering quantum phases with superconducting qubits.
	\end{abstract}

	\maketitle
	
	Analog quantum simulation offers a promising avenue for surpassing classical simulations in certain tasks due to their efficient simulation speeds and minimal resource consumption~\cite{Daley2022,Shaw2024,Nguyen2024,Busnaina2024}. Compared to digital circuits~\cite{Acharya2023,Xu2024,Kim2023,Zhang2022a,Mi2024}, analog quantum simulations exhibit accelerated entanglement growth rates, facilitating the handling of large-scale systems without the need for deep gate circuits~\cite{Xu2018,Xu2020,Shi2023}.
	Topologically ordered matter, distinguished by its long-range entanglement, are of particular interest for analog quantum simulation~\cite{Satzinger2021,Scholl2021,Qiao2024} as they exhibit a variety of interesting properties such as emergent gauge fields~\cite{Sachdev2018}, quantum error-correcting codes~\cite{Kitaev2003}, and anyonic excitations with nontrivial statistics~\cite{Nayak2008}. However, these systems are difficult to analyze theoretically and simulate on classical computers, necessitating further experimental investigation to fully understand their properties and dynamics.
	As synthetic quantum systems progress with improved coherence and increased size, enabling deeper temporal evolution of larger systems, quantum simulation becomes prospective to tackle  properties and dynamics of these systems that are otherwise intractable with conventional methods. 
	Recent progress in superconducting qubits~\cite{Satzinger2021}, ultracold atoms~\cite{Schweizer2019} and Rydberg atom arrays~\cite{Bluvstein2022} has revealed signatures of $\mathbb{Z}_2$ Abelian topological order, marking a significant step forward.
	Non-Abelian topological phases present even more complexity and allure~\cite{Wen1991,Nayak2008,Bartolomei2020}, where the exchange of quasiparticles in these systems can lead to non-Abelian unitary operations that act on degenerate ground states, encoding protected quantum information~\cite{Satzinger2021}. This makes non-Abelian anyons not only fascinating from a condensed matter perspective but also crucial for topological quantum computation.
	Very recently, non-Abelian statistics has been observed  by simulating the projective Ising anyons in the toric code
	model~\cite{Bombin2010} with superconducting qubits~\cite{GoogleQuantumAI2023,Xu2024} and creating the ground state wavefunction of non-Abelian $D_4$ topological order with trapped ions~\cite{Iqbal2024}.
	A more prominent example of such systems is the Kitaev model on a two-dimensional honeycomb lattice~\cite{Kitaev2006anyon}.
	Characterized by exchange interactions between neighboring sites described by $\XX$, $\YY$, and $\ZZ$ terms, this model is exactly solvable and features non-Abelian anyonic excitations with fractional statistics. The anisotropic exchange interactions suppress long-range spin order, fostering a quantum spin liquid state, one of the most exotic and sought-after many-body states in condensed matter physics~\cite{Semeghini2021}.
	Despite its theoretical appeal, the Kitaev model is neither readily realized in materials nor easily simulated with synthetic quantum systems.
	Floquet engineering is a promising approach to implement such an intriguing theoretical model~\cite{Kalinowski2023}. It manipulates the effective Hamiltonian by periodically driving the system with an external field, enabling the realization of complex Hamiltonians that are otherwise inaccessible with the native physical system, facilitating the exploration of exotic phases of matter~\cite{Geier2021,Miller2024,Wang2024}.
	
	\begin{figure}[t]
		\begin{center}
			\includegraphics[width=0.45\textwidth]{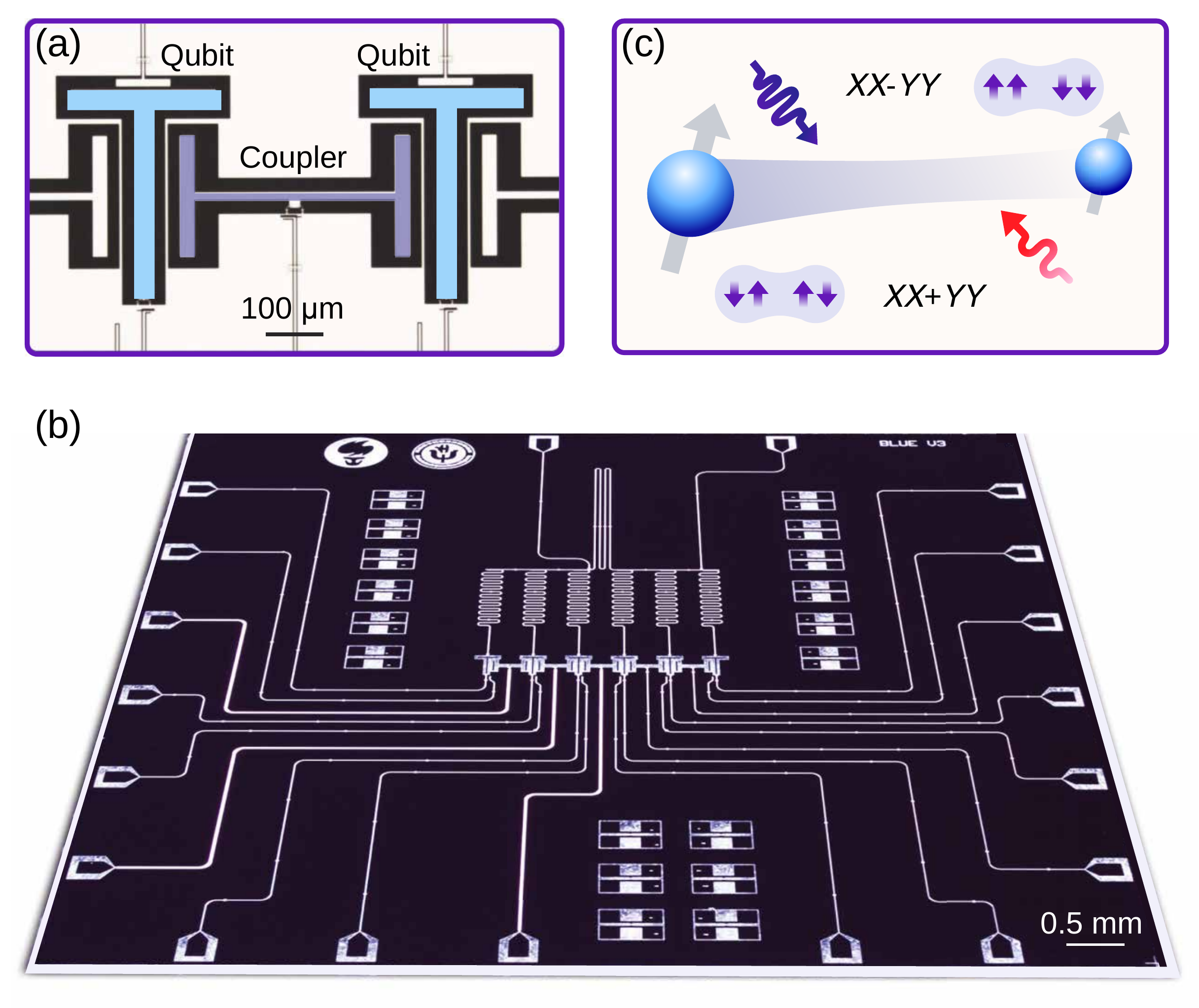}
			\caption{
				\label{fig1}
				Device and schematic diagram.
				(a-b) Photograph of the superconducting quantum processor consisting of six qubits. Panel (a) shows a zoomed-in view of two qubits and a coupler.
				(c) schematic diagram illustrating the blue/red-sideband driving effect.
			}
		\end{center}
	\end{figure}
	
	Superconducting quantum circuits, in particular transmon qubits, have established as a leading candidate for quantum computation~\cite{arute2019quantum}, as well as a flexible platform for quantum simulation, successfully simulating numerous spin models with tens to hundreds of qubits~\cite{Xu2018,Xu2020,Li2023,Shi2023,Kim2023,Zhang2022a,Zhao2022,Roushan2017,Gong2021,GoogleQuantumAI2023,Mi2024}.
	However, physical coupling between transmon qubits naturally yield Jaynes-Cummings Hamiltonians with homogeneous transverse interactions of the form $\XX+\YY$ after rotating wave approximation~\cite{McKay2016}.
	The prohibition of two-photon transitions (counter-rotating terms)~\cite{Niskanen2007,Poletto2012} in such systems limits control over the anisotropy of transverse interactions.
	Several experiments have demonstrated that Floquet engineering can revive two-photon transitions between qubits by applying blue sideband drives to the system, enabling entangling gates~\cite{Leek2010,McKay2016,Roth2017,Nguyen2024empowering,Zhang2024} and autonomous quantum error correction~\cite{Li2024}.
	Floquet engineering also enables full control over the Heisenberg interactions ($\XX$, $\YY$ and $\ZZ$) between two qubits~\cite{Nguyen2024}, showcasing that the native Hamiltonian of superconducting qubits can be transformed into an effective target Hamiltonian.

	Here we present a scalable strategy for generating and calibrating anisotropic transverse interactions with a one-dimensional array of six transmon qubits $Q_i$ ($i=$1 to 6) connected by tunable couplers~\cite{Collodo2020}, as depicted in \rfigs{fig1}(a) and (b). Detailed information regarding the device, experimental setup, and numerical simulation can be found in Ref.~\cite{SM2024}.
	Through Floquet engineering~\cite{Zhao2022,Nguyen2024,Wang2024}, as depicted in \rfig{fig1}(c), we successfully implement pairing ($\XX-\YY$) and hopping ($\XX+\YY$) interactions and achieve independent control over the $\XX$ and $\YY$ terms.
	Observed Aharonov-Bohm interference~\cite{Roushan2016,Roushan2017,Cohen2019,Martinez2023} in synthetic space confirms the tunable and coherent nature of these Floquet engineered interactions.
	To further validate its scalability and applicability in quantum simulation, we demonstrate the transverse field Ising chain (TFIC) model and its dynamical phase transition from paramagnetic to ferromagnetic phases under varying external field. 
	The scalable synthesis of anisotropic transverse interactions enriches the toolbox for engineering quantum phases with superconducting qubits, paving the way for the implementation of more intriguing models requiring spatially dependent interactions (e.g. Kitaev model) and the  exploration of topological phases with non-Abelian excitations.

	\begin{figure*}[ht]
		\begin{center}
			\includegraphics[width=0.95\linewidth]{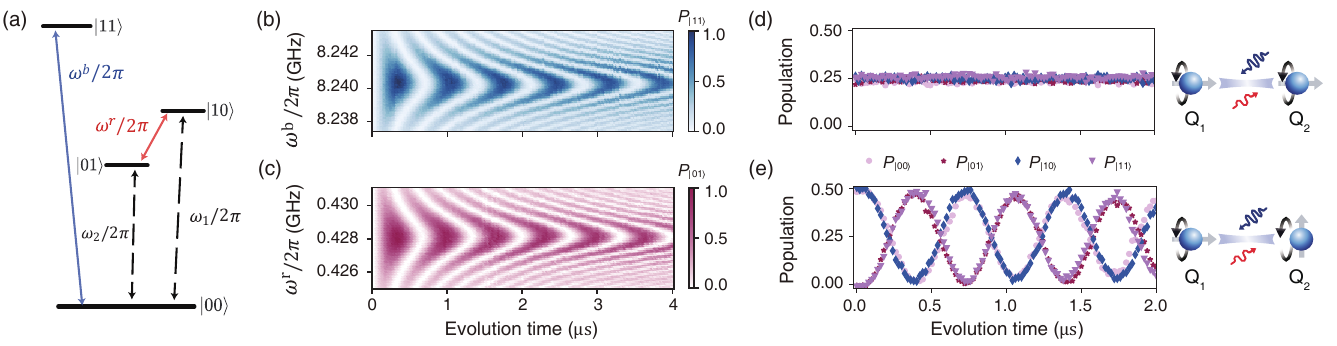}
			\caption{
				\label{fig2}
				Manipulation of pairing ($\XX-\YY$) and hopping ($\XX+\YY$) interactions.
				(a) Energy level diagram of the qubit dimer.
				(b-c) Vacuum Rabi oscillations, $|00\rangle \leftrightarrow |11\rangle$ (top) and $|01\rangle \leftrightarrow |10\rangle$ (bottom), activated by blue and red sideband drives, respectively. The populations of the states $|11\rangle$ ($P_{|11\rangle}$) and $|01\rangle$ ($P_{|01\rangle}$) are shown.
				(d-e) Evolution of the qubit pair under $\XX$ interaction.
				When both qubits are aligned to the $X$ direction in the Bloch sphere, the state remains unchanged under $\XX$ interaction.
				When $Q_2$ is flipped to the $Z$ direction, $\XX$ interaction forces it to rotate around $X$ axis.
			}
		\end{center}
	\end{figure*}

	To elucidate the experimental realization of pairing and hopping interactions, we first focus on a qubit pair, $Q_1$-$Q_2$.
	The  native Hamiltonian of this qubit pair is given by (assuming $\hbar=1$): 
	\begin{shrinkeq}{-1ex}
		\begin{equation}\label{H_2q}
			H=\sum_{i=1}^2\omega_i\sigma_i^+\sigma_i^- + g(t) (\sigma_{1}^+ + \sigma_{1}^-)(\sigma_{2}^+ + \sigma_{2}^-), 
		\end{equation}
	\end{shrinkeq}
	where $\omega_i/2\pi$ and $\sigma_i^+$ ($\sigma_i^-$) denote the frequency and raising (lowering) operator of qubit $Q_i$, respectively. The coupling strength $g(t)$ between qubits can be modulated by the coupler flux~\cite{Roushan2016, Collodo2020, Li2024}.
	When $g \ll \omega_i$, under rotating wave approximation, the coupling Hamiltonian is reduced to hopping (rotating) interactions $\sigma_{1}^+\sigma_{2}^- + \sigma_{1}^-\sigma_{2}^+$ (or $\XX+\YY$), while the pairing (counter-rotating) term $\sigma_{1}^+\sigma_{2}^+ +\sigma_{1}^-\sigma_{2}^-$ (or $\XX-\YY$), is forbidden due to particle number conservation.
	In our experimental implementation, we enable pairing and hopping interactions by applying blue and red sideband drives to the coupler.
	The energy level diagram of the system is depicted in \rfig{fig2}(a), where $|0\rangle$ and $|1\rangle$ are the ground and first excited states of a transmon, $\omega_1/2\pi = 4.336$ GHz, $\omega_2/2\pi = 3.907$ GHz are the transition frequencies of qubit $Q_1$ and $Q_2$ respectively.
	The blue driving frequency is set at $\omega^b/2\pi=(\omega_1+\omega_2)/2\pi=8.240$~GHz to activate the two-photon $\left|00\right\rangle \leftrightarrow \left|11\right\rangle$ transition, and the red sideband driving frequency is set at $\omega^r/2\pi=(\omega_1-\omega_2)/2\pi=0.428$~GHz to activate the $\left|01\right\rangle \leftrightarrow \left|10\right\rangle$ transition respectively.
	By combining the blue and red sideband drives with a frequency diplexer, we achieve concurrent blue and red sideband transitions,  yielding an effective coupling Hamiltonian~\cite{SM2024}
	\begin{eqnarray}\label{H_para}	
		H_{c}=g^b_{1,2} e^{-i\phi^b_{1,2}}\sigma_{1}^{+}\sigma_{2}^{+}+g^r_{1,2} e^{-i\phi^r_{1,2}}\sigma_{1}^{+}\sigma_{2}^{-}+h.c.
	\end{eqnarray}
	Here $g^b_{1,2}$ and $\phi^b_{1,2}$ ($g^r_{1,2}$ and $\phi^r_{1,2}$) are the amplitude and phase of the pairing (hopping) interaction determined by the amplitude and phase of the blue (red) sideband drive respectively.
	when $g^b_{1,2}$ and $g^r_{1,2}$ are set to the same strength, through phase modulation, one could create $\XX$ interaction by choosing $\phi^b_{1,2} = \phi^r_{1,2}$; or create $\YY$ interaction by choosing $\phi^b_{1,2} = \phi^r_{1,2}+\pi$.
	When the qubit pair is initialized at $\left|00\right\rangle$, under pairing interaction, coherent oscillations between $\left|00\right\rangle$ and $\left|11\right\rangle$ are observed, as shown in \rfig{fig2}(b).
	Note the populations are denoted as $P_{|j\rangle}$, where $|j\rangle$ represents the binary representation of each state.
	Similarly, when the qubit pair is initialized at $\left|10\right\rangle$, the hopping interaction induces coherent oscillations between $\left|01\right\rangle$ and $\left|10\right\rangle$ (\rfig{fig2}(c)). Both $g^b_{1,2}/2\pi$ and $g^r_{1,2}/2\pi$ are modulated to 0.75~MHz here.
	By choosing $\phi^b_{1,2} = \phi^r_{1,2}$, the concurrent pairing and hopping interactions yield an $\XX$ interaction.
	When both qubits are aligned to the $X$ direction in the Bloch sphere and subjected to this $\XX$ interaction, the state remains unchanged throughout the evolution. Such a characteristic feature of this interaction is clearly seen in \rfig{fig2}(d) where we prepare the qubit pair in the state $(\left|0\right\rangle+\left|1\right\rangle)\otimes(\left|0\right\rangle+\left|1\right\rangle)/2$ by rotating both qubits to the $X$ direction, and observe that the qubit populations are unchanged with evolution time.
	Conversely, when $Q_2$ is flipped to the $Z$ direction, i.e., the qubit pair is initialized in the state $(\left|0\right\rangle+\left|1\right\rangle)\otimes\left|0\right\rangle/\sqrt{2}$, then the $\XX$ interaction forces $Q_2$ to rotate around the $X$ axis in the Bloch sphere while leaving $Q_1$ unchanged, as illustrated in \rfig{fig2}(e).
	These results demonstrate the dynamic consistency and controllability of the interaction between qubits with Floquet engineering, providing clear evidence for the anisotropy nature of the transverse interaction.

\begin{figure}[t]
		\begin{center}
			\includegraphics[width=0.45\textwidth]{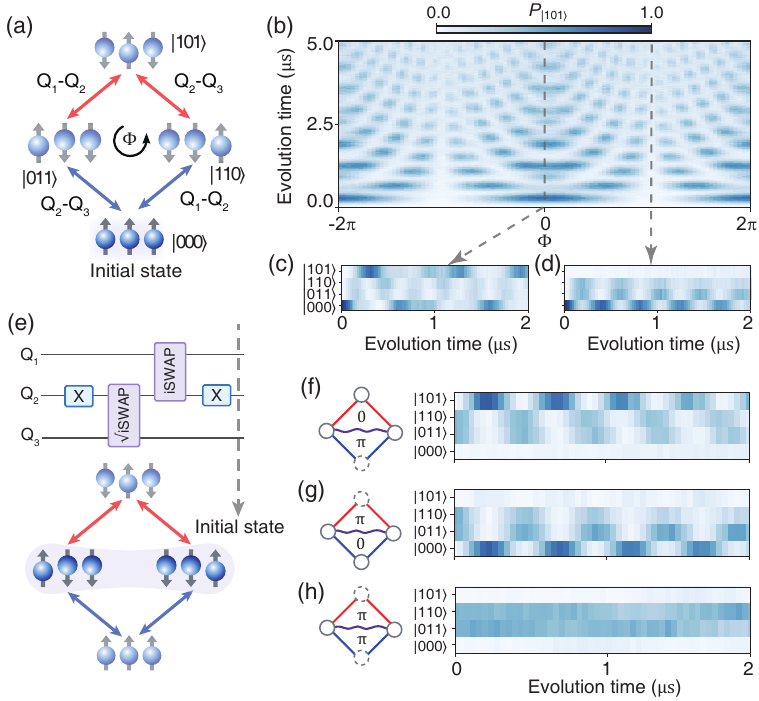}
			\caption{
				\label{fig3}
				Parametric phase dynamics and Aharonov-Bohm (AB) interference.
				(a) Schematic of a qubit trimer illustrating a closed loop with gauge-invariant phase $\Phi$, involving adjacent pairing and hopping interactions within the subspace $\{|000\rangle, |110\rangle, |011\rangle, |101\rangle\}$.
				(b) AB interference as $\Phi$ varies, represented by the population of the state $|101\rangle$ ($P_{|101\rangle}$).
				(c-d) Evolution of state populations across all four states for $\Phi=0$ and $\Phi=\pm\pi$.
				(e) Circuit diagram for preparing the entangled initial state $(|110\rangle+|011\rangle)/\sqrt{2}$.
				(f), (g) Modulation of blue (red) sideband driving phase between $Q_1$ and $Q_2$ induces AB interference, shown here for $\Phi = \pi$.
				(h) Evolution of four states under simultaneous red and blue sideband manipulation at $\phi=\pi$.
			}
		\end{center}
	\end{figure}

	As the number of qubits increases, more intriguing phenomena emerge from the phase tunability of the interactions.
	For example, when considering three qubits connected in a chain, a closed-loop structure emerges in the synthetic space of the qubit basis states $\{\left |000\right \rangle,\left |110\right \rangle,\left |011\right \rangle,\left |101\right \rangle\}$, as shown in \rfig{fig3}(a), where red arrows represent hopping interactions activated by red sideband drives, blue arrows denote pairing interactions activated by blue sideband drives respectively. 
	Within this loop, a gauge-invariant geometrical phase
	$\Phi = \phi_{1,2}^r + \phi_{2,3}^b - \phi_{2,3}^r - \phi_{1,2}^b$ arises. 
	Adjusting the blue/red sideband phases ($\phi_{i,i+1}^{b,r}$, $i=1,2$) modulates the enclosed flux $\Phi$, allowing for the demonstration of an analogy of the Aharonov-Bohm interference effect~\cite{Aharonov1959}.
	By adjusting each of these four phase terms, one can observe diverse AB interference patterns, see~\cite{SM2024} for details.
	Here we show a concrete example in \rfig{fig3}(b), where the initial state $\left|000\right\rangle$ is subjected to four equal-strength pairing/hopping interactions. The system evolves along two distinct paths: $\left|000\right\rangle \to \left|110\right\rangle \to \left|101\right\rangle$ and $\left|000\right\rangle \to \left|011\right\rangle \to \left|101\right\rangle$. By setting $\phi_{2,3}^b$, $\phi_{1,2}^r$, and $\phi_{2,3}^r$ to zero, modulation of the loop flux $\Phi$ is achieved through the change of $\phi_{1,2}^b$. The evolution patterns of the four states with varying $\Phi$ are observed, exemplified by the behavior of the state $\left|101\right\rangle$ (see \rfig{fig3}(b)); see~\cite{SM2024} for the evolution of other states. At $\Phi=0$, all four states exhibit vacuum Rabi oscillations (see~\rfig{fig3}(c)). However, at $\Phi=\pi$, destructive interference significantly affects the state $\left|101\right\rangle$, suppressing its population (see~\rfig{fig3}(d)). This phenomenon, known as Aharonov-Bohm caging~\cite{Vidal1998}, confines the population to other states while impeding occupation of $\left|101\right\rangle$.
	To further elucidate our system's phase tunability, we demonstrate the independent and collaborative modulation of blue and red sideband driving phases. Now we prepare the system in the entangled state $(|110\rangle + |011\rangle)/\sqrt{2}$ as shown in \rfig{fig3}(e); see~\cite{SM2024} for this entangled state preparation and characterization. Starting from $\left|000\right\rangle$, destructive interference consistently occurs at the state $\left|101\right\rangle$, regardless of which sideband phase is modulated. However, starting from this entangled state, destructive interference shifts to different locations depending on the phase manipulation of the respective sideband drives, with interference effects depending on the flux modulation within the top and bottom triangular loops divided by the entangled state. For clarity, we present the population of different states under specific phases, with detailed phase modulation results provided in Ref.~\cite{SM2024}.
	In \rfig{fig3}(f), $\phi_{1,2}^b$ is tuned to $\pi$, while the other three phase terms are set to 0. With this phase configuration, destructive interference is observed at $\left|000\right\rangle$.
	Similarly, in \rfig{fig3}(g), $\phi_{1,2}^r$ is tuned to $\pi$, while the other phase terms are 0.
	With this phase configuration, destructive interference is observed at $\left|101\right\rangle$.
	In a more intricate scenario as shown in \rfig{fig3}(h), all the four phase terms $\phi_{i,i+1}^{b,r}$ ($i=1,2$) are manipulated, where $\phi_{1,2}^{b} =\phi_{2,3}^{r} = \pi/2$ and $\phi_{1,2}^{r} =\phi_{2,3}^{b} = -\pi/2$. This results in a magnetic flux of $\pi$ in both the top and bottom triangle loops, inducing destructive interference and localizing the state in its initial position.

	\begin{figure*}[ht]
		\begin{center}
			\includegraphics[width=0.95\textwidth]{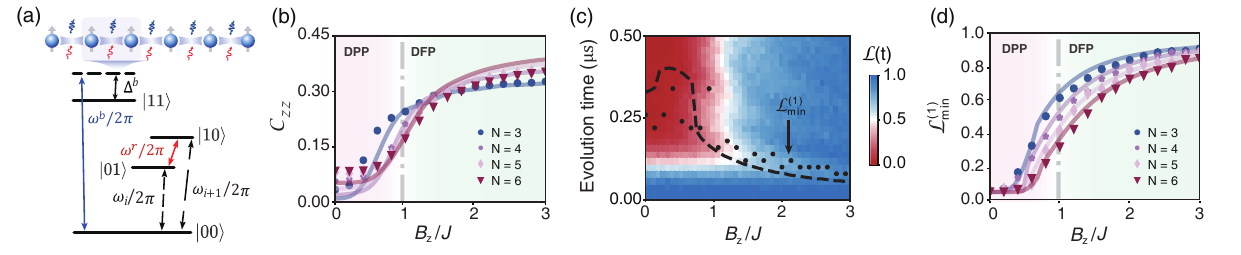}
			\caption{
				\label{fig4}
				Dynamical phase transition of the TFIC.
				(a) Schematic diagram illustrating the construction of a TFIC using detuned blue sideband driving.
				(b) Experimental data (markers) and numerical results (lines) showing the averaged spin correlation $\mathcal{C}_{zz}$ as a function of normalized magnetic field strength $B_z/J$. 
				(c) Dynamics of the Loschmidt echo $\mathcal{L}(t)$ as a function of time 
				and $B_z/J$ for $N=6$. The highlighted black dots are the first minimum of $\mathcal{L}(t)$, while the dashed line is numerical simulation.
				(d) Variation of the first minimum of $\mathcal{L}(t)$ with $B_z/J$, with experimental data shown by markers and numerical results by lines.
			}
		\end{center}
	\end{figure*}

	Continuous modulation of both the amplitude and phase of interaction is crucial for studying the dynamics of quantum many-body systems~\cite{Zhang2017,Bernien2017,Harris2018,Jepsen2020}.
	Such capability has been successfully demonstrated with concurrent blue and red sideband drives in \rfig{fig3}.
	This approach can be extended to larger qubit systems to investigate the transverse field Ising chain ~\cite{Calabrese2011,Heyl2013,Bojan2018,Mi2024} and its dynamical phase transitions (DPT)~\cite{Zhang2017,Xu2020}.
	TFIC have been extensively explored with superconducting quantum annealers~\cite{King2023}, where the dynamics were observed at a time scale much larger than the qubit coherence time.
	Digital as well as Floquet approach that approximate the transverse Ising Hamiltonian have also been demonstrated with programmable superconducting quantum processors~\cite{Dupont2022,Mi2022,Mi2024}.
	Here we demonstrate a coherent synthesis of the TFIC Hamiltonian with six superconducting qubits.
	Building upon \req{H_para}, we extend our system to six qubits ($N=6$) and apply concurrent blue and red sideband drives to the couplers.
	The modulation phases of the sideband drives are carefully calibrated through the AB interference effect as demonstrated in \rfig{fig3}(b) so that the transverse interactions are aligned to the same direction ($\XX$ to be specific, see \cite{SM2024} for details).
	To create an artificial external field for the TFIC model,
	the blue sideband drive frequencies are slightly detuned from the $\left|00\right\rangle \leftrightarrow \left|11\right\rangle$ transition, i.e., $\omega_{i,i+1}^{b} = \omega_{i}+\omega_{i+1} + \Delta^{b}$, as illustrated in \rfig{fig4}(a). 
	With this detune, the concurrent sideband drives yield an effective Hamiltonian of
	\begin{shrinkeq}{-1ex}
		\begin{seqnarray}\label{H_eff}
			\begin{split}
				H_{\mathrm{eff}} = \sum_{i=1}^N\frac{\Delta^b}{4}\sigma_i^z +\sum_{{i=1}}^{{N-1}}( & g_{i,i+1}^be^{-i\phi_{i,i+1}^b}\sigma_i^+\sigma_{i+1}^+ +\\
				&g_{i,i+1}^re^{-i\phi_{i,i+1}^r}\sigma_i^+\sigma_{i+1}+h.c.).
			\end{split}
		\end{seqnarray}\noindent
	\end{shrinkeq}
	Note that next-nearest neighboring qubits are slightly detuned to avoid crosstalk between them.
	By setting $g_{i,i+1}^{b}=g_{i,i+1}^{r}=J = 2\pi\times 0.75$~MHz and $\phi_{i,i+1}^b=\phi_{i,i+1}^r=0$ as calibrated by AB interference, our system can be mapped to the standard TFIC model:
	\begin{sequation} \label{H_Ising}
		H_{\text{Ising}} = \sum_{i=1}^{N-1}J\sigma_i^x\sigma_{i+1}^x+\sum_{i=1}^NB_z\sigma_i^z,
	\end{sequation}\noindent
	where $B_z=\Delta^b/4$ is the external field.
	To demonstrate the capability of our superconducting qubit chain in capturing the dynamical phase transition of TFIC, we vary the normalized magnetic field strength $B_z/J$ by changing the detuning $\Delta^b$ and measure the average-time spin correlation and Loschmidt echo following the system's evolution under \req{H_Ising} for a duration of time $t$.
	First, we conduct measurements of the average-time spin correlation function defined as 
	$\mathcal{C}_{zz}\equiv\frac{1}{T}\int_0^T\sum_{i,j;i\neq j}\langle\sigma_i^z\sigma_j^z\rangle/N^2dt$~\cite{Xu2020,Zhang2017}, here $T=500$ ns is the final evolution time.
	\rfig{fig4}(b) shows $\mathcal{C}_{zz}$ versus different normalized magnetic field strength $B_z/J$ for qubit chain sizes $N=3$, 4, 5 and 6. For $B_z/J<1$, $\mathcal{C}_{zz}$ exhibits relatively small values, indicating weak spin correlations characteristic of the dynamical paramagnetic phase (DPP). Conversely, for $B_z/J>1$, $\mathcal{C}_{zz}$ increases, signaling the onset of strong spin correlations in the dynamical ferromagnetic phase (DFP). Near $B_z/J \approx 1$, $\mathcal{C}_{zz}$ reaches a local minimum, marking the transition to the DFP. This transition becomes more pronounced as the system size $N$ increases~\cite{Bojan2018}.
	Additionally, we validate the transition between the  DPP and the  DFP using Loschmidt echo measurements~\cite{Bojan2018}, which quantifies the overlap amplitude between the initial quantum state and its time-evolved state:
	$\mathcal{L}(t) = |\langle\psi(0)|e^{-iH_{\text{Ising}}t}|\psi(0)\rangle|^2$, here $|\psi(0)\rangle=|0\rangle^{\otimes N}$ is the initial state.
	Non-analytic behavior in the rate function $\mathcal{R}(t) = -(1/N)\log|\mathcal{L}(t)|$ signals the presence of Lee-Yang-Fisher zeros~\cite{Heyl2013, Bojan2018} associated with the DPP-DFP transition as $\mathcal{L}(t)$ approaches zero. Analyzing the minima of $\mathcal{L}(t)$ provides critical insight into phase transitions~\cite{Bojan2018}.
	\rfig{fig4}(c) illustrates the evolution of the Loschmidt echo with evolution time and $B_z/J$,
	specifically for the $N=6$ case (more data for $N=3$, 4 and 5 are available in Ref.~\cite{SM2024}). 
	From this, we extract the first minimum of the Loschmidt echo ($\mathcal{L}^{(1)}_{\text{min}}$), with \rfig{fig4}(d) presenting its dependence on normalized magnetic field strength across different system sizes ($N=3$, 4, 5, 6).
	Initially converging to zero, $\mathcal{L}^{(1)}_{\text{min}}$ significantly increases with $B_z/J$, indicating the transition to the DFP.
	Given the relatively small size of our system, we are not able to observe a sharp transition at the critical point $B_z/J = 1$. Nonetheless, comparative analysis between \rfig{fig4}(b) and \rfig{fig4}(d) reveals different conceptual frameworks for elucidating the TFIC dynamical phase transition, demonstrating a high degree of overlap with numerical calculations based on the generic TFIC model. This consistency confirms the successful manipulation of the TFIC's dynamical phase transition using our one-dimensional superconducting qubit chain.

	In conclusion, with Floquet engineering, we have  achieved concurrent pairing and hopping interactions in transmon qubits, whose superposition yields independent and coherent control over the anisotropy of the transverse interaction.
	Furthermore, we demonstrate the scalability of this approach and realize the transverse field Ising model in a one-dimensional array of transmon qubits, exploring its dynamical phase transition behavior from the paramagnetic phase to the ferromagnetic phase under varying magnetic field.
	Our demonstration paves the way for the implementation of more complex physical systems requiring spatially dependent interactions, enriching the toolbox for engineering quantum phases with superconducting qubits.
	Looking forward, extending this capability to two-dimensional qubit arrays assembled with flip-chip technologies~\cite{arute2019quantum} seems straightforward, except for the considerable engineering challenge. Such an extension would allow for the construction and exploration of quantum compass model~\cite{Nussinov2015} which  has remarkable symmetry properties and the ground state shows topological degeneracy~\cite{FernandezLorenzo2016}.
	Utilizing ancillary energy levels, $\ZZ$ interactions could also be engineered~\cite{Collodo2020,Nguyen2024},  opening a pathway to the implementation of Kitaev model on honeycomb lattices~\cite{Kitaev2001,Busnaina2024}.

	\vspace{15pt}
	\textbf{Data availability}
	
	The data that support the plots within this paper and
	other findings of this study are available from the corresponding
	author upon reasonable request.
	\vspace{15pt}
	
	\begin{acknowledgments}
		We thank Jianjian Miao and Zi Cai for helpful discussions. 
		This work was supported by the National Natural Science Foundation of China (12174178 and 12374474), the Innovation Program for Quantum Science and Technology (2021ZD0301703), the Guangdong Provincial Key Laboratory (2019B121203002), the Science, Technology and Innovation Commission of Shenzhen Municipality (KQTD20210811090049034), the Shenzhen-Hong Kong Cooperation Zone for Technology and Innovation (HZQB-KCZYB-2020050), the Innovation Program for Quantum Science and Technology (2021ZD0301703) and the Guangdong Basic and Applied Basic Research Foundation (2022A1515110615, 2024A1515011714).
	\end{acknowledgments}
	
\bibliography{Main}		
	
\end{document}


\title{Supplementary Material for ``Floquet Engineering of Anisotropic Transverse Interactions in Superconducting Qubits''}
	
	\maketitle
	
	\setcounter{equation}{0}
	\setcounter{figure}{0}
	\setcounter{table}{0}
	
	\renewcommand{\theequation}{S\arabic{equation}}
	\renewcommand{\thefigure}{S\arabic{figure}}
	\renewcommand{\thetable}{S\arabic{table}}
	
	\renewcommand{\figurename}{Fig.}
	\def\figureautorefname{Fig.}
	
	\tableofcontents
	\newpage
	
	\section{Experimental Setup}
	\begin{figure}[htp]
		\centering
		\includegraphics[width=1\textwidth]{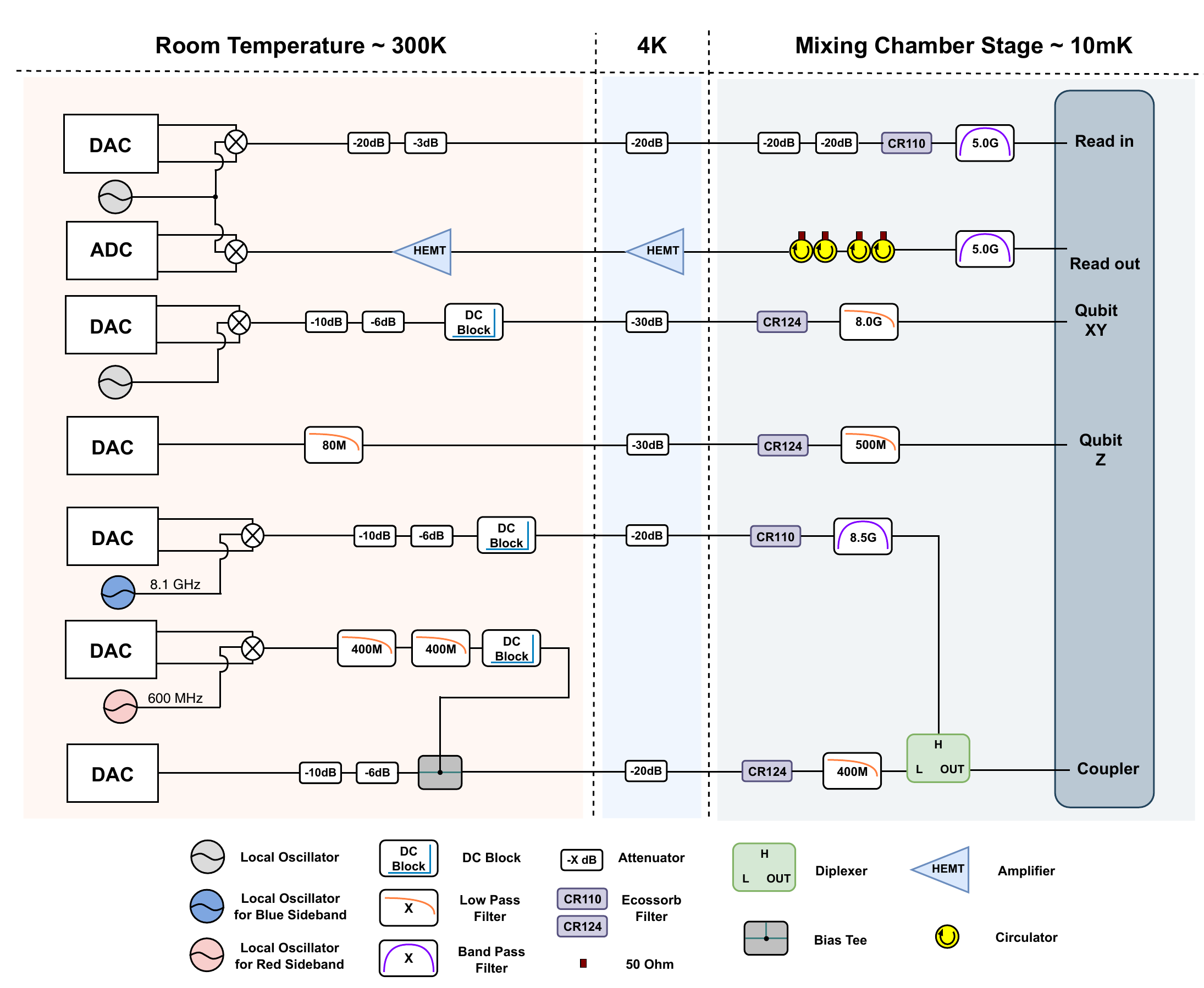}
		\caption{Room temperature and cryogenic wiring.
		}\label{wiring}
	\end{figure}
	
	Figure \ref{wiring} illustrates the room temperature electronics and cryogenic wiring configuration in this experiment. Custom-made digital-to-analog converter (DAC) and analog-to-digital converter (ADC) boards are used for control and measurement of the qubits. 
	The DAC board, with two output channels at 1Gs/s sampling rate, generates direct baseband control signals, or produces microwave control signals via IQ mixing after up-conversion. The baseband signals regulate Z-bias for qubits or the coupler flux, while the microwave signals support XY control and qubit read-in signal, respectively.
	To mitigate thermal noise on the lines, attenuators and filters are integrated in both the XY control lines and the Z bias lines for the qubits.
	The wiring for the couplers are somewhat more involved in this experiment, because both the red ($\sim430$~MHz) and blue ($\sim8.24$~GHz) sideband drive frequencies are beyond the bandwidth of the DAC boards ($\sim250$~MHz). We use dedicated IQ mixers to up-convert signals for the red and blue sideband drives.
	The red sideband drive, generated by down-converting a 600~MHz local oscillator (LO) signal, is filtered through two 400~MHz low-pass filters to suppress LO leakage and the image sideband. 
	To mitigate noise on the blue sideband driving line, an 8.5~GHz bandpass filter with a 600~MHz bandwidth is added in line. Furthermore, to suppress magnetic flux noise, DC blocks are employed on both the blue sideband driving line and the red sideband driving line. The DC bias for the coupler is instead combined with the red sideband signal using a Bias Tee. Finally, after individual filtering, the blue and red driving lines are combined within the mixing chamber stage of the dilution refrigerator using a diplexer.
	
	The read-in line undergoes proper attenuation and filtering at each temperature stage to achieve optimal read-in power while minimizing the impact of thermal noise on qubit coherence. 5~GHz bandpass filters are used to effectively suppress noise outside the bandwidth of the readout resonators. At the 4~K stage, the readout signal is amplified by a high electron mobility transistor (HEMT) low noise amplifier from Low Noise Factory, then further amplified by another HEMT amplifier at room temperature. Subsequently, the signal is down-converted by an IQ mixer, and then digitized and demodulated by the ADC board to discriminate the qubit states.

	
	\subsection{Device fabrication}
\begin{figure}[h]
		\centering
		\includegraphics[width=0.6\textwidth]{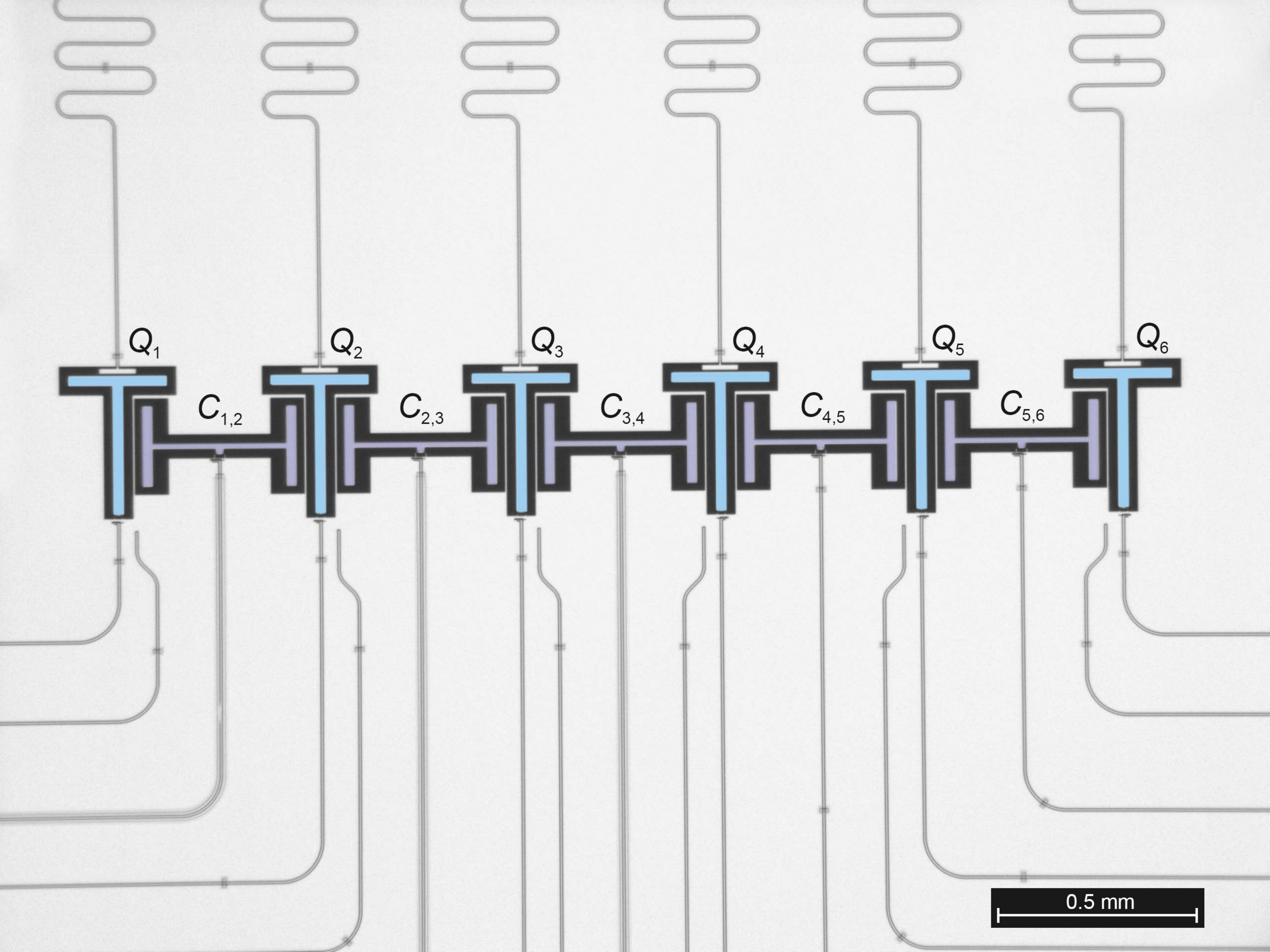}
		\caption{Photograph of the superconducting quantum processor comprising a 1D array of six qubits. The six T-shaped qubits (light blue) are interconnected through H-shaped tunable couplers (light purple).
		}\label{device}
	\end{figure}

	The fabrication process of the superconducting quantum processor involves several steps as briefly described below. First, a 100~nm aluminum film is deposited onto a 2-inch c-plane sapphire wafer using the Plassys MEB system. Photo-lithography, facilitated by laser direct writing, is followed by inductively coupled plasma (ICP) dry etching to define key components such as capacitor pads, readout resonators, control and readout circuits. Subsequently, another round of photo-lithography and electron beam evaporation of 200~nm SiO$_2$ is conducted to create the essential scaffold layer for crossovers~\cite{Dunsworth2018}. The electron-beam lithography (EBL) process is employed next, followed by double-angle evaporation of aluminum utilizing the Plassys system, and lift-off processes, to produce the Al/AlO$_\textrm{x}$/Al Josephson junctions. Further steps involve additional photo-lithography and electron beam evaporation of 300~nm aluminum, along with lift-off processes, to fabricate the conducting layer required for crossovers and bandage layer~\cite{Dunsworth2017} for Josephson junction. \emph{In situ} ion milling process is used for ensuring the establishment of galvanic contact with the bottom aluminum layer formed during the initial deposition. Finally, the wafer is diced into individual dies, each representing a fully functional unit of the superconducting quantum processor as shown in Fig.~\ref{device}.
	
	\subsection{Device performance}

	\begin{table}[htbp]
		\centering
		\label{Tab01}
		\resizebox{0.9\linewidth}{!}{
			\begin{tabular}{p{3.6cm}<{\centering}p{1.2cm}<{\centering}p{1.2cm}<{\centering}p{1.2cm}<{\centering}p{1.2cm}<{\centering}p{1.2cm}<{\centering}p{1.2cm}<{\centering}p{1.2cm}<{\centering}p{1.2cm}<{\centering}p{1.2cm}<{\centering}p{1.2cm}<{\centering}p{1.2cm}<{\centering}}
				\toprule
				component & $Q_1$ & $Q_2$ & $Q_3$ & $Q_4$ & $Q_5$ & $Q_6$ &$C_{1,2}$ &$C_{2,3}$ &$C_{3,4}$ &$C_{4,5}$ &$C_{5,6}$\\
				\midrule
				$\omega^{\textrm{min}}/2\pi$ (GHz)& 3.556 & 3.690 & 3.544 & 3.690 & 3.521 & 3.835\\
				$\omega^{\textrm{max}}/2\pi$ (GHz) & 4.391 & 4.477 & 4.352 & 4.510 & 4.344 & 4.497 & 6.42 & 6.41 & 6.48 & 6.43 & 6.31\\
				$\omega^{\textrm{idle}}/2\pi$ (GHz) & 4.121 & 4.477 & 4.088 & 4.496 & 4.096 & 4.482 &5.44 &5.25 &5.55 & 5.65 &5.39\\
				$\omega^{\textrm{rr}}/2\pi$ (GHz) & 4.957 & 5.067 & 4.984 & 5.095 & 5.024 & 5.135\\
				$E_c/2\pi$ (MHz) & $-$251 & $-$242 & $-$250 & $-$238 & $-$251 & $-$232\\
				$T_1$ ($\mu$s) & 49.8 & 34.8 & 24.1 & 23.0 & 36.8 & 37.2\\
				$T_2^r$ ($\mu$s) & 2.6 & 44.1 & 2.2 & 9.4 & 1.9 & 1.7\\
				$T_2^e$ ($\mu$s) & 17.4 & 41.2 & 29.0 & 33.2 & 19.1 & 37.0\\
				$F_0$ & 0.956 & 0.959 & 0.971 & 0.989 & 0.984 & 0.964\\
				$F_1$ & 0.904 & 0.947 & 0.908 & 0.944 & 0.926 & 0.909\\
				\bottomrule
			\end{tabular}
		}
		\caption{{\bf Device characteristics.} $\omega^{\textrm{min}}/2\pi$, $\omega^{\textrm{max}}/2\pi$ and $\omega^{\textrm{idle}}/2\pi$  represent the minimum, maximum and idle frequencies of the qubits and the couplers respectively. 
			$\omega^\textrm{rr}/2\pi$ are the corresponding readout resonator frequencies. $E_c/2\pi$ represents the anharmonicity of $Q_i$. $T_1$, $T_2^r$ and $T_2^e$ denote the lifetime, Ramsey dephasing time and Spin echo dephasing time of $Q_i$, respectively, which are measured at the idle frequecy. Lastly, $F_0$ ($F_1$) denotes the fidelity of measuring $Q_i$ in the $|0\rangle$ ($|1\rangle$) state when it is initially prepared in the $|0\rangle$ ($|1\rangle$) state.}
	\end{table}

	In this experiment, both the qubits and couplers are frequency-tunable. As illustrated in \rfig{device}, the T-shaped qubits (light blue) are interconnected by H-shaped couplers (light purple). 
	The qubits are fabricated with asymmetric junctions, exhibiting a maximum frequency of approximately 4.5 GHz and a minimum frequency of approximately 3.6 GHz, as illustrated  in \rtab{Tab01}. The average anharmonicity of the qubits is $-240$~MHz. The qubit idle frequencies are categorized into two groups: qubits with odd (even) indices idle near 4.1 GHz (4.5 GHz). The detuning of the qubit frequencies reaches approximately 350 MHz, which is significantly larger than the qubit anharmonicity, thereby mitigating high-level leakage during sideband driving.
	
	The six readout resonators are coupled to a Purcell filter~\cite{Purcellfilter,Purcellfilter1}, which functions as a bandpass filter that facilitates the transmission at the readout resonator frequencies while impeding signals at the qubit frequencies, thus suppressing the Purcell decay rate. The frequency range of the readout resonators spans from 4.9 to 5.1 GHz, well aligned with the center frequency of the Purcell filter. The average fidelity for state $|0\rangle$ during readout is 0.97, while for state $|1\rangle$, it is 0.92. The average qubit energy relaxation time, denoted as $T_1$, is approximately 34~$\mu$s. Apart from $Q_2$, which operates at its maximum frequency, the average dephasing time $T_2^r$ is approximately 2~$\mu$s. Meanwhile, the $T_2^e$ measured by the spin echo sequence is around 27~$\mu$s. In our experiment, the performance of the sideband drive is predominantly affected  by $T_2^r$ and $T_2^e$.
	
	The couplers are essentially transmon qubits with symmetric junctions. The maximum frequencies of the five couplers are approximately 6.4~GHz. When blue and red sideband drives are applied, the couplers are biased to a frequency range of 5.3-5.6~GHz.

	\section{Floquet engineering of anisotropic transverse interaction}\label{Sec.II}

	On the quantum processor, the tunable coupling between qubits is facilitated by a transmon qubit acting as a coupler~\cite{Collodo2020}. Consider the simple case of two qubits $Q_1$ and $Q_2$ coupled with the coupler $C_{1,2}$, the Hamiltonian is expressed as (assuming $\hbar = 1$):
	\begin{equation}
		H_{qc} = \sum_{i=1}^{2} \omega_{i} \sigma_{i}^+ \sigma_{i}^- + \omega_c(t) \sigma_c^+ \sigma_c^-  + \sum_{i=1}^{2} g_{i,c} \left( \sigma_i^+  + \sigma_i^- \right)\left(\sigma_c^+ +  \sigma_c^- \right) + J_{1,2} \left( \sigma_1^+ + \sigma_1^- \right)\left( \sigma_2^+ + \sigma_2^- \right).
	\end{equation}
	Here $\omega_c/2\pi$ is the frequency of the coupler, $g_{i,c}/2\pi \approx 100$~MHz represent the direct coupling strengths between the qubits and the coupler, $J_{1,2}/2\pi \approx 6$~MHz is the direct coupling between the qubits,  $\sigma_i^+$  and $\sigma_i^-$ ($i=1, 2, c$) are the raising and lowering operators for the qubits and the coupler, respectively.
	As the neighboring qubits are largely detuned in this experiment, $|\omega_1-\omega_2|\gg J_{1,2}$, the last term can be ignored here.
	The frequency of the coupler, $\omega_c$, can be tuned via a time-dependent magnetic flux $\Phi_c$ threading its junction loop:
	\begin{equation}
		\omega_c(\Phi_c) = \omega_c^\text{max} \sqrt{\left|\cos\left(\frac{\pi \Phi_c}{\Phi_0}\right)\right|},
	\end{equation}
	where $\Phi_0=2.067\times 10^{-15}$~Wb is the flux quantum, and $\omega_c^\text{max}/2\pi \approx 6.4$~GHz denotes the maximum frequency of the coupler (see Table~\ref{Tab01} for details of each coupler).
	
	\begin{figure}[h]
		\centering
		\includegraphics[width=0.9\textwidth]{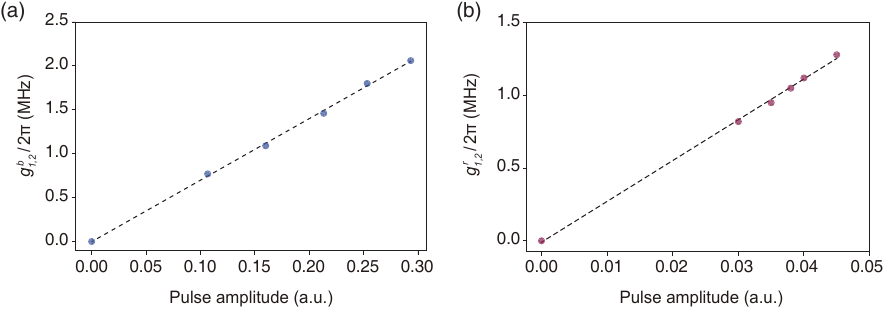}
		\caption{
			Pairing and hopping interaction strengths. (a) The pairing interaction strength ($g^b_{1,2}$) as a function of modulation amplitude. The blue circles represent experimental data, while the black dashed line indicates a linear fit. (b) The hopping interaction strength ($g^r_{1,2}$) as a function of modulation pulse amplitude, with the experimental data shown as red circles.
		}\label{amp_to_g}
	\end{figure}
	
	Experimentally, the coupler frequency is modulated by applying sinusoidal pulses to the flux bias line of the coupler~\cite{Collodo2020, Wang2024,Li2024}, i.e., $\Phi_c(t) = \Phi_{dc}+\Omega_d^b \cos(\omega_d^b t +\phi^b) + \Omega_d^r \cos(\omega_d^r t +\phi^r)$,
	where $\Phi_{dc}$ is the static flux that determines the idle frequency of the coupler, $\Omega_d^b$, $\omega_d^b$ and $\phi^b$ ($\Omega_d^r$, $\omega_d^r$ and $\phi^r$) represent the amplitude, frequency and phase of the blue (red) sideband drive, respectively.
	within the range of suitable experimental parameters, say $\Phi_{dc}$ is properly chosen so that the coupler idle frequency $\omega_c/2\pi$ is in a linear regime and is far detuned from the qubit idle frequency $\omega_i/2\pi$;
	the blue and red sideband drives satisfy
	$\omega_d^b \approx \omega_1+\omega_2$,
	$\omega_d^r \approx \omega_1-\omega_2$,
	and $\Omega_d^{b,r}\ll \Phi_{0}$;
	then this yields an effective Hamiltonian:
	\begin{equation}
		H_{c}\approx g_{1,2}^b e^{-i\phi_{1,2}^b}\sigma_{1}^{+}\sigma_{2}^{+}+g_{1,2}^r e^{-i\phi_{1,2}^r}\sigma_{1}^{+}\sigma_{2}^{-}+h.c.
	\end{equation}
	Here, $g_{1,2}^{b(r)}$ and $\phi_{1,2}^{b(r)}$ represent the strengths and phases associated with pairing (hopping) interactions, respectively, where the strength is given by~\cite{Roth2017}:
	\begin{align}
		& g^b_{1,2}\approx\Omega_d^b\frac{g_{1,c} g_{2,c}}{4}\left.\frac{\partial\omega_c}{\partial\Phi_c}\right|_{\Phi_c=\Phi_{dc}}\left(\frac{1}{(\omega_c-\omega_1)(\omega_c+\omega_2)}+\frac{1}{(\omega_c+\omega_1)(\omega_c-\omega_2)}\right),
		\\
		& g^r_{1,2} \approx \Omega_d^r\frac{g_{1,c}g_{2,c}}{4}\left.\frac{\partial\omega_c}{\partial\Phi_c}\right|_{\Phi_c=\Phi_{dc}}\left(\frac{1}{(\omega_c-\omega_1)(\omega_c-\omega_2)}+\frac{1}{(\omega_c+\omega_1)(\omega_c+\omega_2)}\right).
	\end{align}
	The interaction strengths exhibit a linear relationship with the strength of the external flux.
	But to achieve the desired same coupling strengths, the pulse amplitude of blue sideband drive is approximately one order of magnitude larger than the red sideband drive due to the frequency disparity.
	As illustrated in \rfig{amp_to_g} (a) and (b), the experimental results for both pairing and hopping interactions align well with theoretical predictions~\cite{Roth2017}.
	Note that when the drive strength becomes too strong, the system's coherence degrades, especially for the blue sdieband drive. 
	To mitigate the discrepancy between the strength of  blue and red sideband drive, we adopt separate filtering of the blue and red sideband drives, subsequently merging them using a diplexer (see \rfig{wiring}).
	Given the necessity of simultaneously engaging both hopping and pairing interactions while maintaining the coherence, we ultimately chose an interaction strength of 0.75 MHz for our experiments.

	\begin{figure}[h]
		\centering
		\includegraphics[width=0.9\textwidth]{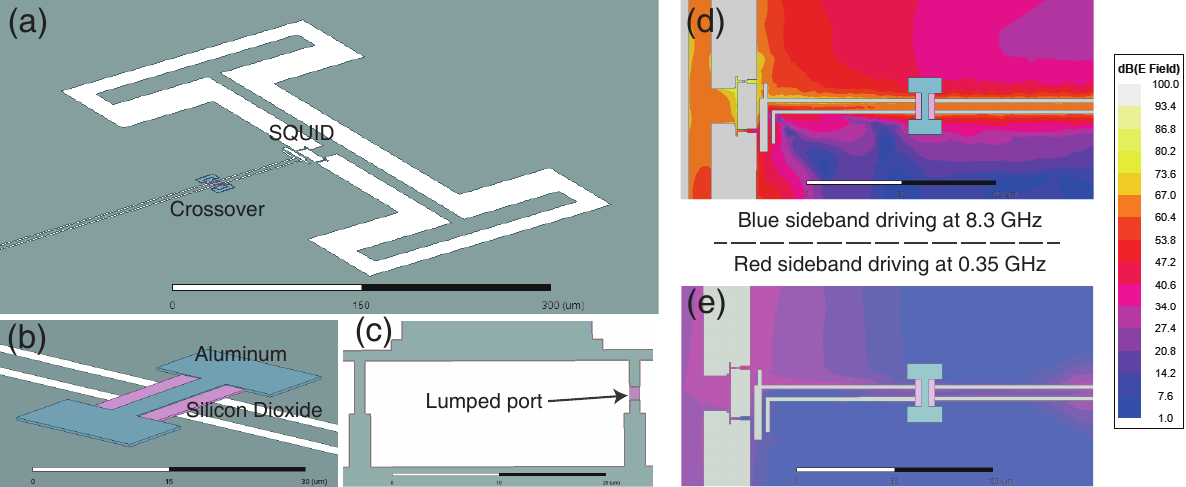}
		\caption{Electromagnetic field simulation of the coupler.
			(a) Design of the coupler capacitance circuit. The green area represents the aluminum (Al) layer, while the white area indicates the bare sapphire substrate.
			(b) Crossover structure in the simulation. The crossover consists of silicon dioxide with an aluminum coating on top. Incorporating a crossover that connects the ground on both sides of the transmission line more closely approximates actual conditions.
			(c) Geometric design of the junction loop. The small pink rectangle represents the lumped port used in the simulation.
			(d) Electric field amplitude distribution at a driving frequency of 8.3 GHz.
			(e) Electric field amplitude distribution at a driving frequency of 0.35 GHz.
		}\label{HFSS_blue_red}
	\end{figure}
	
	In this experiment, the frequency and drive strength for blue sideband is an order of magnitude larger than the red sideband drive.
	As depicted in \rfig{HFSS_blue_red}, numerical simulation results reveal a significant difference in electric field intensity between blue and red sideband driving signals.
	The specific coupler flux bias line is adopted here to  minimize stray inductive coupling to the coupler junction loop to below 1~pH for both blue and red sideband drives.

	\section{Aharonov-Bohm (AB) Interference}
	
	\begin{figure}[hb]
		\centering
		\includegraphics[width=0.9\textwidth]{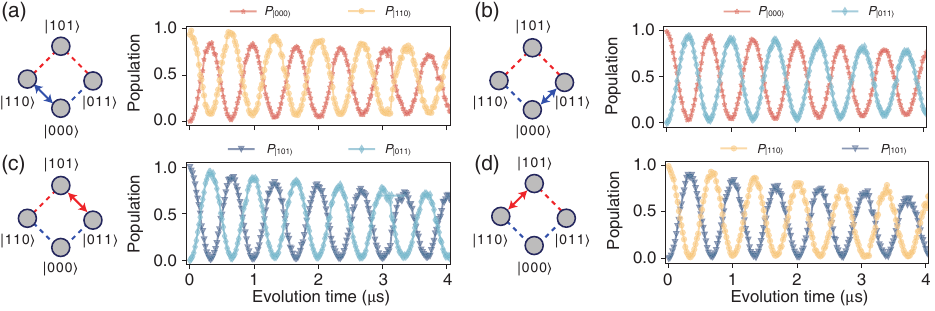}
		\caption{Coherent oscillations in a closed loop formed by pairing and hopping interactions.
			(a) Schematic diagram (left) of a closed loop formed by the states $|000\rangle$, $|011\rangle$, $|101\rangle$, and $|110\rangle$. The blue arrow indicates the pairing interaction between qubits $Q_1$ and $Q_2$. Experimental data (right) shows the population of the state $|000\rangle$ (red dots) and the state $|110\rangle$ (yellow stars).
			(b) The blue arrow signifies the pairing interaction between qubits $Q_2$ and $Q_3$. 
			The cyan rhombuses denote the population of the state $|011\rangle$.
			(c) The red arrow signifies the hopping interaction between qubits $Q_1$ and $Q_2$.
			The blue inverted triangles represent the population of the state $|101\rangle$.
			(d) The red arrow indicates the hopping interaction between qubits $Q_2$ and $Q_3$.
		}
		\label{loop_state}
	\end{figure}
	
	As illustrated in Fig. 3 of the main text, two scenarios with different initial states are presented to observe Aharonov-Bohm (AB) interference. 
	This section provides Supplementary data for Fig. 3 in the main text.

	\subsection{Coherent control of pairing and hopping interactions}

	When the number of qubits reaches three, phase coherence of the Floquet engineered interactions between qubits $Q_1$ and $Q_2$, as well as between $Q_2$ and $Q_3$, must be taken into consideration.
	First we apply blue/red sideband drives individually to make sure their coupling strengths are the same.
	In \rfig{loop_state}, we selectively activate blue/red sideband transitions between each pair of qubits with the same interaction strength of 0.75~MHz.
	This confirms the ability to control and achieve coherent oscillations in a three-qubit chain through pairing and hopping interactions.

	\begin{figure}[htp]
		\centering
		\includegraphics[width=0.9\textwidth]{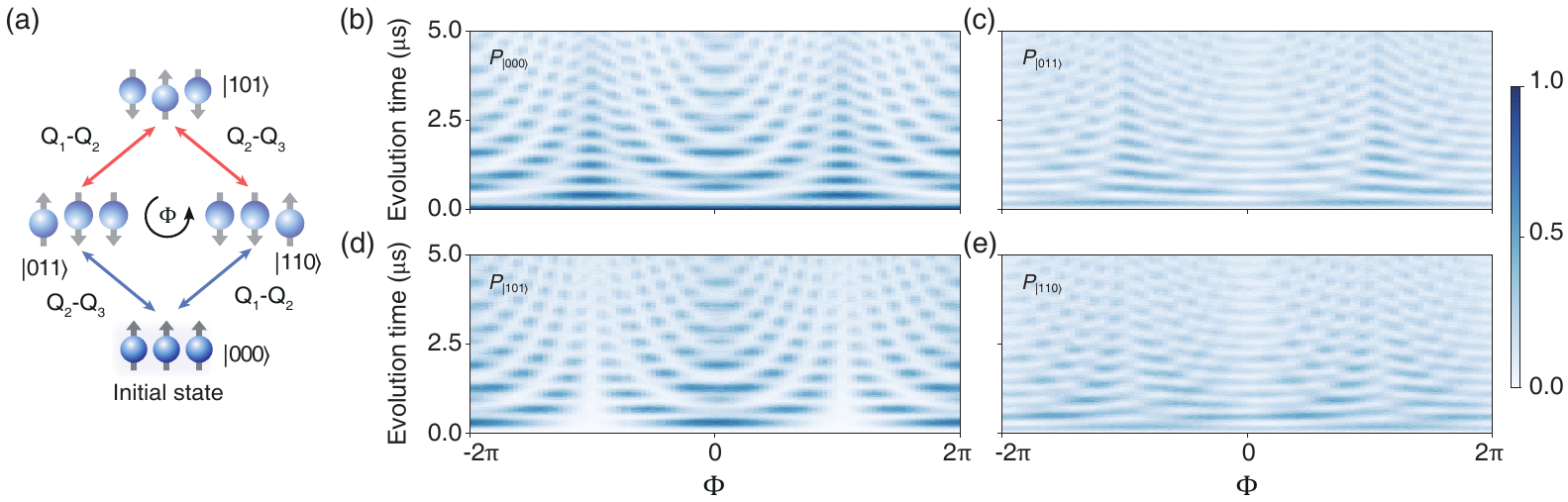}
		\caption{AB  interference in closed-loop states.
			(a) Schematic of a qubit trimer forming a closed loop, illustrating the gauge-invariant geometrical phase $\Phi$.
			(b)-(e) Population dynamics of states $|000\rangle$, $|011\rangle$, $|101\rangle$, and $|110\rangle$ as the flux $\Phi$ varies.
		}
		\label{4site}
	\end{figure}
	
	In \rfig{4site}, we present more AB interference data complementary to Fig.3 in the main text. While Fig.3 in the main text focuses on the population dynamics of the state $|101\rangle$, here we provide the experimental data of all four states within the loop. \rfigs{4site}(b), (c), (d), and (e) respectively depict the detailed population dynamics of the states $|000\rangle$, $|011\rangle$, $|101\rangle$, and $|110\rangle$, highlighting the interference characteristics.

	\subsection{Preparation and calibration of entangled initial state}
	
	\begin{figure}[htp]
		\centering
		\includegraphics[width=0.9\textwidth]{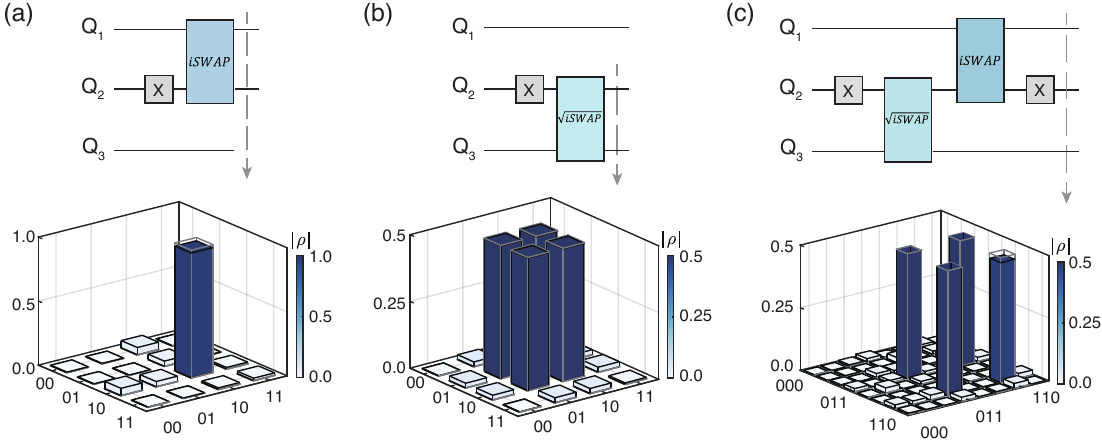}
		\caption{Step-by-step preparation of the entangled initial state.
			(a) Quantum state tomography of $Q_1$ and $Q_2$ after applying an $X$ and an $\sqrt{\rm{iSWAP}}$ gate, with a fidelity of 97.2\%. 
			(b) Applying an $X$ and an $\sqrt{\rm{iSWAP}}$ gate yields a Bell state with a fidelity of 99.0\%. 
			(c) Combining the steps in (a) and (b) yields the desired entangled state $(|110\rangle+|011\rangle)/\sqrt{2}$, with a state fidelity of 96.1\%. 
			Solid bars are experimental data of the density matrix $\rho$ in each panel, whereas gray frames are ideal values.
		}\label{ini_ent_state}
	\end{figure}
	
	To showcase our capability in phase manipulation, we examine a specific scenario depicted in Fig. 3(e) of the main text. The process of preparing an entangled initial state $(|110\rangle+|011\rangle)/\sqrt{2}$ is illustrated in \rfig{ini_ent_state}, with a state fidelity of 96.1\%.

	\subsection{Aharonov-Bohm interference with entangled initial state}
	
	\begin{figure}[!h]
		\centering
		\includegraphics[width=\textwidth]{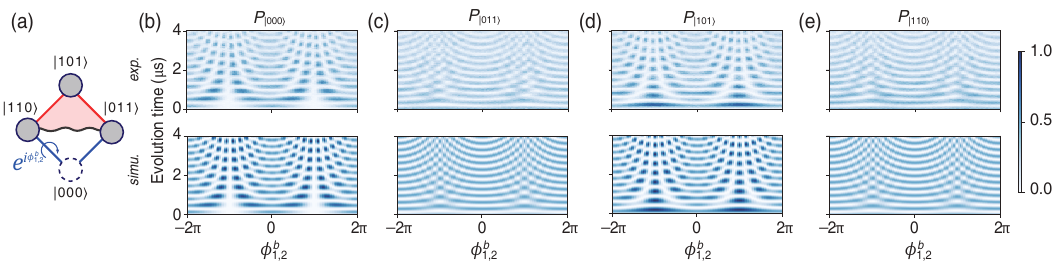}
		\caption{AB interference with varying $\phi_{1,2}^b$.
			(a) Schematic of a closed-loop system with states $|101\rangle$, $|011\rangle$, $|000\rangle$, and $|110\rangle$ arranged clockwise. Blue and red lines represent pairing and hopping interactions. The black wavy line signifies entanglement between $|011\rangle$ and $|110\rangle$. The blue arrow indicates the experimentally controlled phase of the blue sideband drive between qubits $Q_1$ and $Q_2$.
			(b-e) Population distributions of states $|000\rangle$, $|011\rangle$, $|101\rangle$, and $|110\rangle$. The upper panels display experimental data, while the bottom panels show numerical simulation results.
		}\label{ini_ent_b12}
	\end{figure}
	
	Following the preparation of the initial state $(|110\rangle+|011\rangle)/\sqrt{2}$ shown in \rfig{ini_ent_state}(c), we demonstrate the independent and cooperative modulation effects of blue and red sideband drive phases. As detailed in Fig. 3(f-h) of the main text, the entangled state divides the diamond-shaped loop into upper and lower triangular segments. Adjusting the phases of the respective sideband drives induces destructive interference, causing quantum states to transition between different positions. This interference effect crucially depends on the modulation of magnetic flux within these triangular segments.
	
	\begin{figure}[!h]
		\centering
		\includegraphics[width=\textwidth]{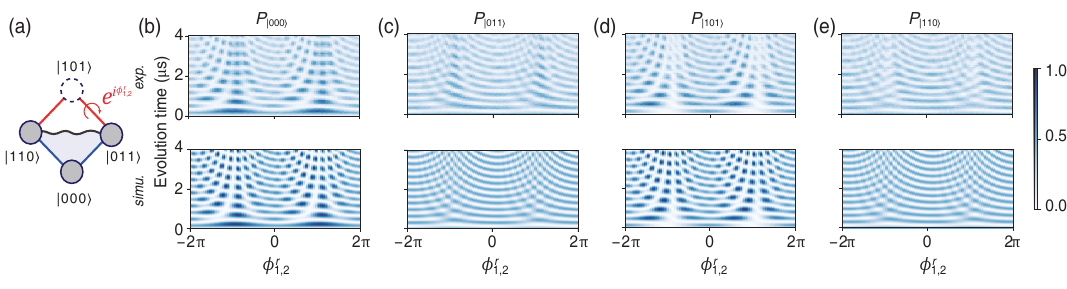}
		\caption{AB interference with varying $\phi_{1,2}^r$.
			(a) Closed-loop system with states $|101\rangle$, $|011\rangle$, $|000\rangle$, and $|110\rangle$. Red arrow shows the phase of the red sideband drive between $Q_1$ and $Q_2$.
			(b-e) Population distributions of $|000\rangle$, $|011\rangle$, $|101\rangle$, and $|110\rangle$. Experimental data (top), numerical simulations (bottom).
		}\label{ini_ent_r12}
	\end{figure}
	
	\begin{figure}[!h]
		\centering
		\includegraphics[width=\textwidth]{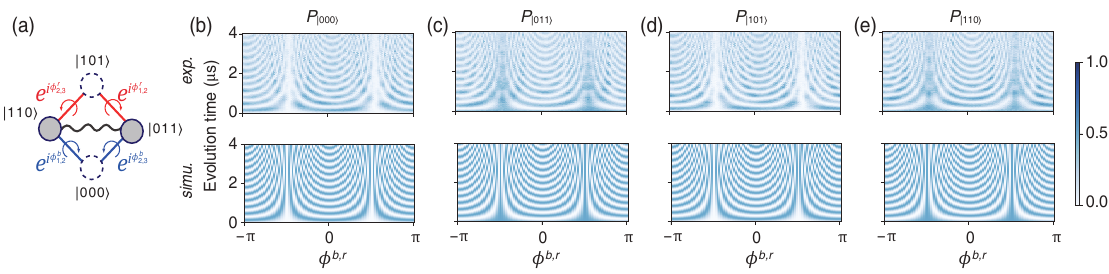}
		\caption{AB interference with varying $\phi_{1,2}^b$ and $\phi_{1,2}^r$.
			(a) Closed-loop system with states $|101\rangle$, $|011\rangle$, $|000\rangle$, and $|110\rangle$. Blue and red arrows indicate the phases of blue and red sideband drives, respectively. The phases satisfy $\phi_{1,2}^b=\phi_{2,3}^r=-\phi_{1,2}^r=-\phi_{2,3}^b=\phi^{b,r}$.
			(b-e) Population distributions of $|000\rangle$, $|011\rangle$, $|101\rangle$, and $|110\rangle$. Experimental data (top), numerical simulations (bottom).
		}\label{ini_ent_br}
	\end{figure}

	We investigate three specific scenarios: firstly, exclusively varying the blue sideband phase $\phi_{1,2}^b$ between qubits $Q_1$ and $Q_2$. \rfig{ini_ent_b12} illustrates the temporal evolution of the four states as $\phi_{1,2}^b$ ranges from $-2\pi$ to $2\pi$. Destructive interference is observed at $|000\rangle$ when $\phi_{1,2}^b = \pm\pi$, with populations distributed among the remaining three states. Similarly, exclusive modulation of the red sideband phase $\phi_{1,2}^r$ between qubits $Q_1$ and $Q_2$ facilitates continuous flux modulation in the upper triangular segment, demonstrating destructive interference at $|101\rangle$ for $\phi_{1,2}^r=\pm\pi$, as depicted in \rfig{ini_ent_r12}.
	
	Finally, by simultaneously adjusting all sideband drive phases $\phi_{i,i+1}^{b,r}$ ($i=1,2$) as shown in \rfig{ini_ent_br}, where $\phi_{i,i+1}^{b,r}$ satisfies $\phi_{1,2}^b=\phi_{2,3}^r=-\phi_{1,2}^r=-\phi_{2,3}^b$, destructive interference is observed at both the upper and lower segments when $\phi_{1,2}^{b,r}=\pi/2$. This configuration corresponds to magnetic fluxes of $\pi$ in both triangular loops, effectively localizing the states back to the initial entangled state.

	\subsection{Phase calibration for TFIC}
	
	\begin{figure}[!h]
		\centering
		\includegraphics[width=0.6\textwidth]{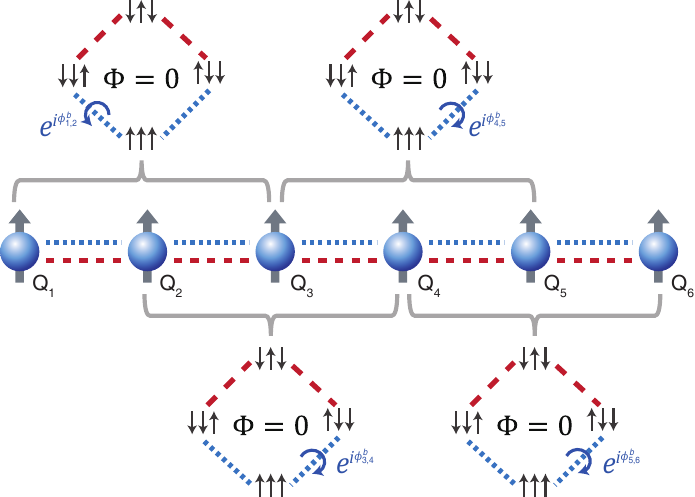}
		\caption{Phase calibration diagram for a six-qubit chain. The blue and red dashed lines denote the pairing and hopping interactions, respectively. During phase calibration, the coupling phases $\phi_{1,2}^b$, $\phi_{3,4}^b$, $\phi_{4,5}^b$, and $\phi_{5,6}^b$ are calibrated for the loops formed by $Q_1$-$Q_2$-$Q_3$, $Q_2$-$Q_3$-$Q_4$, $Q_3$-$Q_4$-$Q_5$, and $Q_4$-$Q_5$-$Q_6$, respectively. By observing the AB interference pattern like \rfig{4site}, we calibrate the net loop flux to zero.
		}\label{phase_cali_simu}
	\end{figure}

	To validate the applicability and scalability of concurrent pairing and hopping interactions, we demonstrate the implementation of the transverse field Ising chain (TFIC) model with six qubits.
	While the synthesis of a six-qubit TFIC is not a feat, the Floquet engineering of anisotropic transverse interactions in a scalable fashion paves the way for the implementation of more complex physical systems such as quantum compass model and the Kitaev model, enriching the toolbox for engineering quantum phases with transmon qubits.
	
	When extending the system beyond three qubits, it is essential to calibrate the phases of the sideband drives.  As illustrated in \rfig{phase_cali_simu}, qubits are depicted as blue spheres, with blue sideband coupling represented by a blue dotted line and red sideband coupling by a red dotted line. Each set of three adjacent qubits is encapsulated within parentheses, forming a unit. The schematic diagram of a synthetic four-site loop, constructed from the states $|000\rangle$, $|011\rangle$, $|101\rangle$, and $|110\rangle$, is simplified in \rfig{phase_cali_simu}, where upward-pointing arrows indicate the state $|0\rangle$ and downward-pointing arrows indicate $|1\rangle$. By adjusting the phase of the blue sideband couplings $\phi_{1,2}^b$, $\phi_{3,4}^b$, $\phi_{4,5}^b$, and $\phi_{5,6}^b$ in each three-qubit unit, we observe the AB interference pattern like \rfig{4site}, from which we calibrate the loop flux to zero. After this calibration, all transverse interactions are aligned to the $X$ direction along the chain.

	\section{Supplementary data for dynamical phase transition}
	
\begin{figure}[ht]
		\centering
		\includegraphics[width=0.9\textwidth]{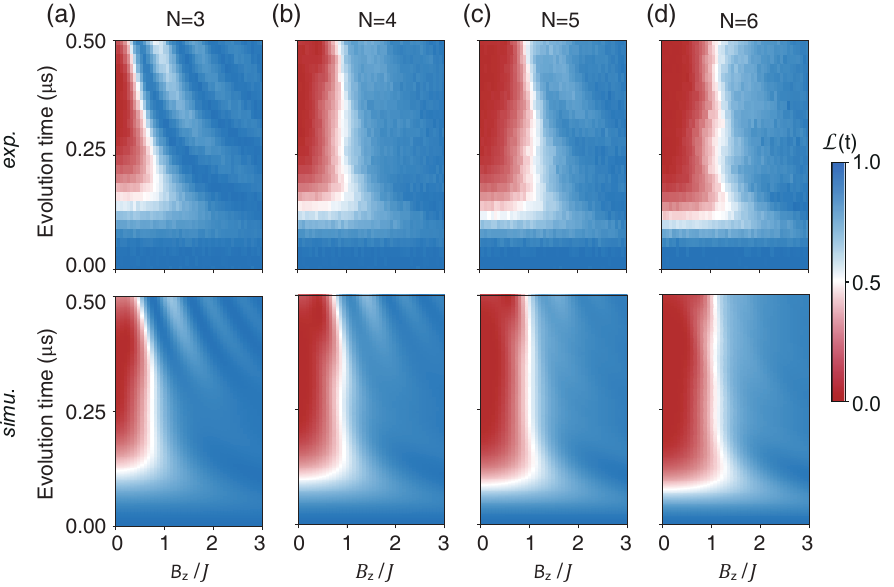}
		\caption{Dynamics of the Loschmidt echo $\mathcal{L}(t)$ across different system sizes and varying magnetic field strength.
			Results for (a) three-qubit, (b) four-qubit, (c) five-qubit, and (d) six-qubit chain systems. Experimental data are presented in the upper panel, while numerical simulations are shown in the bottom. $B_z/J$ represents the normalized strength of the transverse field.
		}
		\label{LEN3N6}
	\end{figure}

	The Loschmidt echo, given by $\mathcal{L}(t) = |\langle\psi(0)|e^{-iH_{\text{Ising}}t}|\psi(0)\rangle|^2$, represents the overlap amplitude between the initial state $|\psi(0)\rangle=|0\rangle^{\otimes N}$ and its time-evolved state.
	In Fig.4(c) of the main text, we only showed the data for $N=6$ qubits.
	In \rfig{LEN3N6}, we present experimental data alongside numerical simulations of the Loschmidt echo $\mathcal{L}(t)$ across various system sizes ($N=3$, 4, 5 and 6 respectively), as a function of time and the normalized magnetic field strength $B_z/J$. Here, the field strength is defined as $B_z=\Delta^b/4$, with a coupling strength of $J/2\pi=0.75~  \text{MHz}$.
	
	When $B_z/J$ is small, the system resides in the dynamical paramagnetic phase (DPP), during which $\mathcal{L}(t)$ decreases rapidly toward zero. As $B_z/J$ increases, the system transitions to the dynamical ferromagnetic phase (DFP), where $\mathcal{L}(t)$ remains non-zero in the absence of decoherence effects. At the point where $\mathcal{L}(t)=0$, the rate function $\mathcal{R}(t)=-(1/N)\log|\mathcal{L}(t)|$ becomes non-analytic, signaling the presence of Lee-Yang-Fisher zeros associated with the DPP-DFP transition~\cite{Bojan2018}. We focus on the first minimum of the Loschmidt echo, denoted as $\mathcal{L}_{\rm{min}}^{(1)}$, which allows us to investigate behavior at short evolution times. By examining $\mathcal{L}_{\rm{min}}^{(1)}$ as a function of $B_z/J$, we can observe the phase transition between DPP and DFP through changes in $\mathcal{L}_{\rm{min}}^{(1)}$ as $B_z/J$ varies.

	\clearpage